\documentstyle[12pt,psfig,aaspp4]{article}

\begin{document}

\title{Using Perturbative Least Action to Reconstruct Redshift Space
Distortions}

\author{David M. Goldberg}
\authoremail{goldberg@yale.princeton.edu}
\affil{Yale University, Astronomy Dept.,  New Haven, CT, 06520-8101}
\affil{Princeton University Observatory, Princeton, NJ 08544-1001}

\begin{abstract}
In this paper, we present a redshift space reconstruction scheme which
is analogous to and extends the Perturbative Least Action (PLA) method
described by Goldberg \& Spergel (2000).  We first show that this
scheme is effective in reconstructing even nonlinear observations.  We
then suggest that by varying the cosmology to minimize the quadrupole
moment of a reconstructed density field, it may be possible to lower
the errorbars on the redshift distortion parameter, $\beta$ as well as
to break the degeneracy between the linear bias parameter, $b$, and
$\Omega_M$.  Finally, we discuss how PLA might be applied to realistic
redshift surveys.
\end{abstract}

\section{Introduction}

The current generation of galaxy redshift surveys is producing an
avalanche of data about the structure of the universe.  The Sloan
Digital Survey (SDSS; York et al. 2000) has already measured the
redshifts of $\sim 30,000$ galaxies (X. Fan, private communication) of
the $\sim 10^6$ galaxy redshifts over 10,000 ${\rm deg}^2$ which the
survey will ultimately cover.  The Two Degree Field redshift survey
(2dF; Colless 1999) has measured redshifts for $\sim 10^{5}$ galaxies,
and will eventually measure a quarter of a million galaxies over 2000
${\rm deg}^2$ in the southern hemisphere.  Meanwhile, the IRAS 0.6 Jy
Point Source Catalog redshift Survey (PSCz; Saunders et al. 2000),
with about $15,000$ galaxies over virtually all of the sky, provides a
fertile testbed for cosmological models and methods.

As impressive as these surveys are, they are limited to providing a
somewhat distorted snapshot of the universe.  For example, it is both
suggested observationally (Hubble 1936; Oemler 1974; Davis \& Geller
1976; Kaiser 1984; Santiago \& Strauss 1992; Blanton 2000 and
references therein) and predicted (Davis et al. 1985; Bardeen et
al. 1986; Blanton 1999 and references therein; Dekel \& Lahav 1999)
that the luminous structure, that which the surveys record, is biased
with respect to the underlying matter density of the universe.
Additionally, though the Hubble relation can be used to give an
approximate 3-dimensional picture of structure, peculiar velocities
(e.g. Strauss \& Willick 1995 and references therein for a review)
cause a distortion of the structure along the line of sight.  Since
the peculiar velocity field is a function of the underlying mass field
and cosmology, velocity distortions and bias are intimately related.

Perturbative Least Action (PLA; Goldberg \& Spergel 2000, hereafter
GS) was shown to be an excellent technique for the reconstruction of
nonlinear structure in real space.  In this paper, we extend PLA into
redshift space, and show that it is an excellent tool for extracting
information from redshift surveys even into the nonlinear regime.

However, before going too far afield, it will be useful to review some
of the basic issues involved in redshift space distortions of the
density field, and define some of the symbols which will be used
throughout this paper.  This discussion is not meant to be
comprehensive, however, and the interested reader will certainly
benefit from some of the excellent reviews on the subject (Hamilton
1998; Hatton \& Cole, 1998; Zaroubi \& Hoffman 1996; Strauss \&
Willick 1995; Sahni \& Coles 1995; Kaiser 1987).

\subsection{Definitions and Conventions}

Let us consider an observer in an expanding universe.  Hubble's law
states that in a Friedman-Robertson-Walker universe, the distance, $d$ to a
test particle with redshift, $z$, will be, to first order in $z$:
\begin{equation}
cz\simeq H_0d^{FRW}\ ,
\end{equation}
where $H_0$ is the Hubble constant at the present epoch, $d$ is the
distance to the particle, and the approximation comes from the fact
that relativistic effects become important at high redshifts.
However, we will confine our discussion to the non-relativistic case
and this approximation throughout this discussion.

Hubble's Law assumes that a particle (galaxy) is at rest in comoving
coordinates.  A particle with a local, peculiar velocity, ${\bf v}$,
will have a redshift of:
\begin{equation}
cz=H_0d+{\bf \hat{x}}\cdot {\bf v}\ ,
\label{eq:pecvel}
\end{equation}
where ${\bf \hat{x}}$ points along the line of sight of the test
particle.  Since it is actually this redshift that we observe, and not
the position of the particle, it is worthwhile to construct a comoving
coordinate which reflects the observations of the observer at the
origin.  We define the comoving redshift space coordinate:
\begin{equation}
{\bf s}={\bf x}+{\bf \hat{x}}\left(\frac{a{\bf \hat{x}}\cdot \dot{\bf
x}}{\dot{a}}\right)\ , 
\label{eq:sdef}
\end{equation}
where we have substituted $a\dot{\bf x}$ for ${\bf v}$ and $a{\bf x}$
for $d\hat{\bf x}$ in our implicit definition of the peculiar
velocity. Comparison with equation~(\ref{eq:pecvel}) yields the
relation, $s(t)=cz/\dot{a}(t)$.

Equation~(\ref{eq:sdef}) shows that the mapping of ${\bf x}$ to
${\bf s}$ is inherently non-invertible.  Any trajectory of ${\bf x}$
will yield a single trajectory in redshift space, but the converse
does not necessarily hold.

Indeed, even in the Zel'dovich regime (defined below), the infall of
matter from both sides (front and back) of a structure can give rise
to a ``triple-valued zone'' (see Strauss \& Willick 1995, \S 5 for a
discussion), as matter at different physical distances from the
observer appear to have identical redshifts due to conspiracy between
the Hubble flow and the peculiar velocity.  We will attempt to
disentangle this degeneracy in \S~\ref{sec:method}.

\subsection{The Linear Regime}

For now, let us consider the linear regime.  By convention, particle
$i$ sits at position, ${\bf q}_i$, at $a=0$, and that the ensemble of
${\bf q}_i$ forms a uniform grid. If a field is linear, that is, if
\begin{equation}
\delta({\bf x},t)\equiv \frac{\rho({\bf x},t)}{\overline{\rho}(t)}-1
\label{eq:lin}
\end{equation}
remains small at all times, and if the velocity field initially has no
curl, then the Zel'dovich approximation (Zel'dovich 1970) can be used
to give the trajectory of a particle as:
\begin{equation}
{\bf x}_i(t)={\bf q}_i+D(t){\bf p}_i\ ,
\label{eq:zeld}
\end{equation}
where $D(t)$ is a cosmology dependent, monotonically increasing growth
factor, normalized to unity at the present, and ${\bf p}_i$ is the
final displacement of particle, $i$, from its initial position.

Fields for which equation~(\ref{eq:zeld}) well approximates the
trajectories of all particles at all times will be referred to as
Zel'dovich fields.  This is to be distinguished from fields in the
``linear regime'' for which:
\begin{equation}
\delta({\bf q}_i,t)=-D(t)\nabla_q\cdot {\bf p}({\bf q}_i)
\label{eq:lin}
\end{equation}
holds at all times.  When perturbations are very small, both
equalities hold.  However, as perturbations become larger,
equation~(\ref{eq:lin}) breaks down first, and densities evolve
according to a more complex function of time.

In the Zel'dovich regime, the redshift space coordinate evolves as:
\begin{equation}
{\bf s}(t)^{(zel)}={\bf q}+D(t){\bf p}+\hat{\bf x}\left[
\hat{\bf x}\cdot\left(\frac{a(t)\dot{D}(t)}{\dot{a}(t)}{\bf p}\right)
\right]=
{\bf q}+D(t){\bf p}+\hat{\bf x}\left[D(t)f(\Omega_M,\Omega_\Lambda,t)
\hat{\bf x}\cdot{\bf p}
\right]
\end{equation}
where we define $f(\Omega_M,\Omega_\Lambda,t)$ such that:
\begin{equation}
f(\Omega_M,\Omega_\Lambda,t)\equiv
\frac{a(t)\dot{D}(t)}{\dot{a}(t)D(t)}\ .
\end{equation}
For $\Omega_M=1$, this function is a constant in time.  At
$t=t_0$, a good analytic fit can be given by
$f(\Omega_M,\Omega_\Lambda,t_0)\simeq \Omega_M^{0.6}$ (Peebles 1980).
This function is normally used in discussions of bias, and is
generally combined with the linear bias parameter, $b$, to relate the
divergence of the velocity field to the overdensity of the galaxy
field at the present day, via the parameter, $\beta\equiv
f(\Omega_M,\Omega_\Lambda,t_0)/b$, where $\delta_g=b\delta$ for a
linear biasing model.  

In this section, we will be treating only unbiased fields, and have
introduced $\beta$ as a means of comparing this discussion of redshift
space distortions to the standard approach (e.g. Strauss \& Willick
1995).  In \S3, we'll return to the degeneracy in $\beta$ and show
how PLA might be used to break it.

\subsection{The Distant Observer Approximation}

Up to this point, we've treated linear redshift space distortions with
more or less full generality.  However, since our ultimate goal is to
apply these distortions in the context of PLA, we will want to make
some simplifying assumptions.  For example, PLA (GS) uses a Particle Mesh (PM) Poisson solver (Hockney \& Eastwood
1981).  This method takes advantage of Fast Fourier Transforms (FFTs),
which assume Cartesian coordinates.

The general form of the comoving redshift coordinate, ${\bf s}$, above
(equation~\ref{eq:sdef}), is not separable in Cartesian
coordinates.  If this form were to be applied in general, one would
wish to describe coordinates with spherical harmonics (e.g. Susperregi
2000).  The primary purpose of the current discussion, however, is to
examine the underlying dynamics, and while treating redshift space
distortions of nearby systems is undoubtedly of cosmological interest,
the matter at hand is greatly simplified by assuming the distant
observer approximation (d.o.a.).

In the d.o.a. we essentially assume that the system of interest is
sufficiently far away that the $\hat{\bf x}_i$ are parallel for all
particles.  For convenience, we will label the third orthonormal
coordinate (the z-axis), as the line of sight.  Using this definition,
we redefine the comoving redshift coordinate:
\begin{equation}
s^{\alpha}=
x^{\alpha}+\delta^K_{\alpha3}\frac{a\dot{x}^{\alpha}}{\dot{a}}\ ,
\end{equation}
where $\alpha$ denotes the index of the direction vector, and
$\delta_{\alpha\beta}^K$ is the Kronecker-$\delta$ function.

In the linear regime, this becomes
\begin{equation}
s^{\alpha}(t)=q^{\alpha}+D(t)p^{\alpha}[1+\delta^K_{\alpha
3}f(\Omega_M,\Omega_\Lambda,t)]\ .
\label{eq:sdos}
\end{equation}

Thus, if a particle is observed at ${\bf s}_i^F$ and its initial
position, ${\bf q}_i$, and the cosmology are known, this expression
may be inverted and combined with equation~(\ref{eq:zeld}) to
give:
\begin{equation}
x^\alpha(t)=q^\alpha+D(t)p^{\alpha}=q^\alpha+ \frac{D(t)(
s^{\alpha F}-q^{\alpha})}{[1+\delta^K_{\alpha
3}f(\Omega_M,\Omega_\Lambda,t_0)]} \ .
\label{eq:xdos}
\end{equation}

Equations~(\ref{eq:sdos},\ref{eq:xdos}) may be combined to
extend linear theory into redshift space.  One must keep in mind that
we have assumed that we know both the initial and final position of a
particle in the comoving redshift space coordinate.  As pointed out
above, without both constraints, inverting the redshift coordinate
becomes ill-posed.

\section{Method: Least Action in Redshift Space}
\label{sec:method}

Even if we have the idealized set of observations discussed above, and
have a complete, unbiased mapping of the density field in redshift
space, as perturbations become large, complications will arise both in
mapping the redshift space field to a real space one, and in
reconstructing an initial density field.  A number of researchers have
attempted worked on the dual problems of reconstructing an underlying
real space CDM density field from a velocity field, and the
calculation of a velocity field field from a redshift survey.

For the former problem, the POTENT algorithm (Bertschinger \& Dekel
1989; Dekel, Bertschinger \& Faber 1990; Dekel et al. 1999) uses the
Zel'dovich approximation to relate a redshift/distance survey, in
which one component of the peculiar velocity can be directly computed,
to an underlying mass density field.  Nusser et al. (1991) uses a
nonlinear generalization to extend this into the nonlinear regime.
Others (Kudlicki et al. 1999; Chodorowski et al. 1998; Chodorowski \&
\L okas 1997; Bernardeau 1992) use higher order perturbation theory to
compute the relationship between the velocity divergence and real
space density field.  In particular, Chodorowski \& \L okas (1997)
point out that application of these methods may be used to break the
degeneracy between the linear bias constant, $b$, and cosmology.

A related problem concerns the calculation of the peculiar velocity
field from a galaxy redshift survey.  Nusser \& Davis (1994) use a
quasi-linear correction to the Zel'dovich approximation to relate the
redshift space density field to the peculiar velocity field.  Like the
methods listed above which use perturbation theory to related the
velocity and density field, Chodorowski (2000) uses 2nd- and 3rd-
order perturbation theory to expand and compare the real space and
redshift space density fields.

Others have taken a slightly different approach, which attempts to
essentially solve these two problems simultaneously.  Giavalisco et
al. (1993) suggested that the Least Action approach described by
Peebles (1989) could be used in redshift space with only a canonical
transform of the coordinates.  Schmoldt and Saha (1998) consider the
difficulties of running least action reconstruction in redshift space,
and test this by reconstructing the velocity field of the Local Group.

Susperregi \& Binney (1994) modified this approach and described a
technique whereby one could expand the density and velocity fields in
Fourier space, and write down the Least Action equations in Eulerian
form.  Susperregi (2000) took this a step further, and applied a
similar code (using Spherical Harmonic transforms) to the IRAS 1.2 Jy
redshift survey (Fisher et al. 1995).  Each of these techniques use a
smoothing filter on the density field such that they are in the mildly
nonlinear regime at the present.  

PLA has several distinct advantages over these approaches.  First,
since they are inherently Eulerian, as perturbations becomes large,
they no longer fairly sample the matter field.  PLA, on the other
hand, is Lagrangian in the sense that it performs the time integral
over the particle field, rather than the density field.  Moreover, the
Eulerian PLA approaches assume a locally curl-free velocity field at
all times, by the velocity-density relationship.  While PLA generally
assumes curl-free initial conditions, vorticity is permitted to
develop.

Ultimately, we want to reconstruct the underlying real space CDM
density field and evolution from a set of observed galaxy redshifts.
The approach taken in this paper differs from those discussed in
GS in that we now deal with quasi-linear
structure, rather than the highly nonlinear constraints.  While
previously, we were content to get a ``realistic'' set of initial
conditions, here, our aim is reconstruct the details of the
observations exactly.  By doing this, we hope to disentangle
degeneracies in bias, get a handle on the true cosmological power
spectrum of perturbations, and examine the growth of large scale
structure.

In this section, we describe applying the PLA approach to redshift
space constraints.  In particular, we will deal with two main issues:

\begin{enumerate}
\item Given some observed redshift space density field, $\delta^F({\bf
s})$, find final particle constraints, $\{ {\bf s}^F_i \}$, which
satisfy the density field, which maps to a corresponding initial
uniform field, $\{ {\bf q}_i \}$, such that the constraints can be
most easily satisfied.
\item Given an initial particle constraints, $\{ {\bf q}_i \}$, and final
particle constraints in redshift space, $\{ {\bf s}^F_i \}$, find the
trajectories of particles which self-consistently satisfy the boundary
conditions.
\end{enumerate}

\subsection{Computing Particle Constraints}

We must first determine the boundary conditions for the particles'
positions at $a=0$ and $a=1$.  Let's consider a set of idealized
observations, in which a smoothed, complete, and unbiased density
field in redshift space, $\delta^F({\bf s})$, has been observed.  

In order to determine the final boundary condition, we generate a set
of particle redshift positions which yield this observed density
field.  We begin by assuming by distributing particle positions
uniformly distributed on a grid, $\tilde{\bf s}_i={\bf q}_i$, where
the tilde over the redshift coordinate will be explained shortly.

From here, we iterate in the following way.  In each iteration, we
take the density field of the current value of the particle positions
yielding $\delta({\bf s})$.  We then transform this density field into
the target field by using the Jacobian determinant to create a laminar
flow.  That is, one can adjust the particle positions in the former
grid by using a coordinate transformation:
\begin{equation}
\tilde{\bf s}_i'=\tilde{\bf s}_i+\nabla \psi(\tilde{\bf s}) \ .
\end{equation}
In that case, the density field as measured in the primed frame
compared to the unprimed frame will be:
\begin{equation}
\left [1+\delta({\bf s}')\right] d^3{\bf s}'=\left[1+\delta({\bf
s})\right] d^3{\bf s} \ .
\end{equation}
From the form of the transform, the determinant of the Jacobian is
easily computed.  For small perturbations:
\begin{equation}
\left[1+\delta({\bf s}')\right]\simeq \left[1+\delta({\bf s})\right]
\left(1+\nabla^2\psi \right)\ .
\label{eq:psi_eq}
\end{equation}
Assigning $\delta({\bf s}')=\delta^F({\bf s})$, and using standard
Fourier techniques, one can compute the scalar field, $\psi({\bf s})$,
and taking the gradient, one can compute the coordinate transform, and
therefore, can iteratively produce a particle map which satisfies the
target density field. 

Since we assume throughout that structure evolves out of an
initially uniform density field, the initial constraints on these
particles must be the uniform grid, ${\bf q}_i $.  This is exactly
analogous to mapping from an unperturbed simulation since a uniform
particle field run through an N-body will remain uniform, and hence
serve as a perfectly legitimate unperturbed simulation.

But which final positions correspond to the values of ${\bf q}_i$?  In
other words, given any set of particle redshift positions, $\tilde{\bf
s}_i^F$ which satisfy the density field, $\delta^F({\bf s})$, what is
the ``best'' permutation matrix, ${\bf M}_{ij}$, such that
\begin{equation}
{\bf s}^F_i = \sum_j {\bf}M_{ij}\tilde{\bf s}_j^F\ ?
\end{equation}
We define ${\bf M}_{ij}$ as an $N_p\times N_p$ matrix which has
exactly one ``1'' in each row and column, and zeros elsewhere.

We need to define what we mean by ``best''.  In general, we mean that
we wish to compute boundary constraints which most naturally provide
us with physically well-motivated orbits.  Since the laminar flow
method of generating final constraints necessarily assumes no shell
crossings, for small perturbations, the permutation matrix will simply
be the identity matrix.  In this ultra-linear case, in which
perturbations are assumed to be so small that there are no orbit
crossings even in redshift space, we estimate the physical particle
displacement from the uniform field as:
\begin{equation}
p^{\alpha}_i=\frac{\tilde{s}_i^{\alpha
F}-q_i^{\alpha}}{1+f(\Omega_M,\Omega_\Lambda,t)\delta^K_{\alpha 3}}\ ,
\label{eq:partpos}
\end{equation}
The final physical position of particle, $i$, is thus related by
equation~(\ref{eq:zeld}).  Taking the smoothed density field of
${\bf x}_i$ gives an estimate of the density field in real space.

However, in many cases of interest, even where there are no shell
crossings in real space, there {\it are} orbit crossings in redshift
space.  These are the famous ``triple value zones,'' (Strauss \&
Willick 1995 \S5.9) so named because a particle at a given redshift is
a triply degenerate function of distance.  Even if a system is
dynamically in the linear regime and the real space density field is
known with great accuracy, this degeneracy can arise.

\begin{figure}
\centerline{\psfig{figure=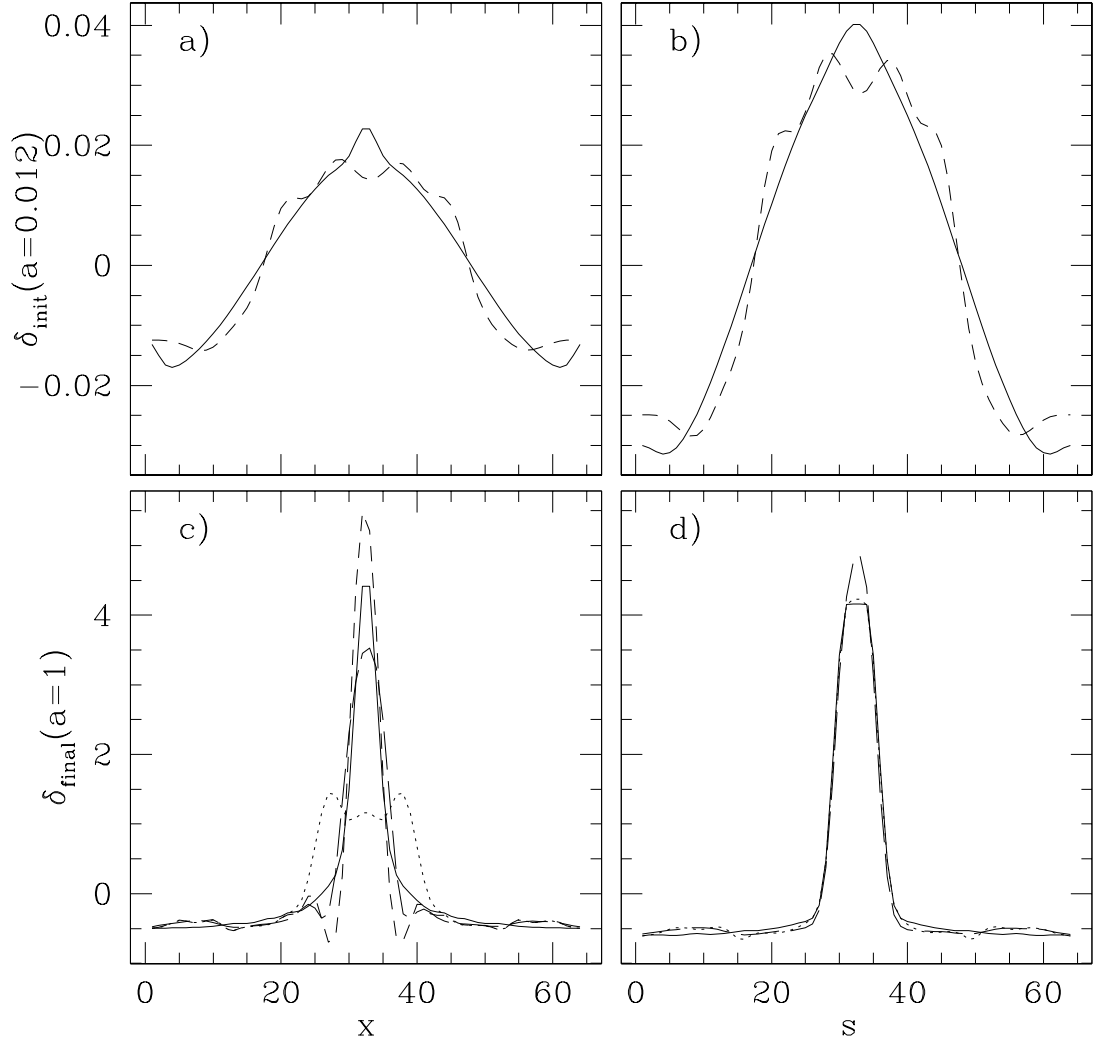,angle=0,height=6in}}
\caption{A simulation of a quasi-linear Zel'dovich pancake.  In each
panel, the solid lines show the ``true'' real- and redshift space
density fields before and after running the corresponding particles
through an N-body code. Panels a) and b) show the initial density
fields in redshift and real space, respectively.  The long-dashed
lines represent the initial density field computed by using PLA.
Panel c) shows the evolved field in real space.  The solid line shows
the true field, while the dotted line shows the first estimate of the
field from the redshift space distribution, and the short dashed line
shows the converged value.  The long dashed line shows the density
field computed using PLA.  Panel d) shows the evolved field in
redshift space.  The dotted line denotes the density field
corresponding to the redshift positions generated by using the laminar
flow density matching method described in the text.  The long dashed
line shows the density field computed by using PLA.}
\label{fg:1d}
\end{figure}

\begin{figure}
\centerline{\psfig{figure=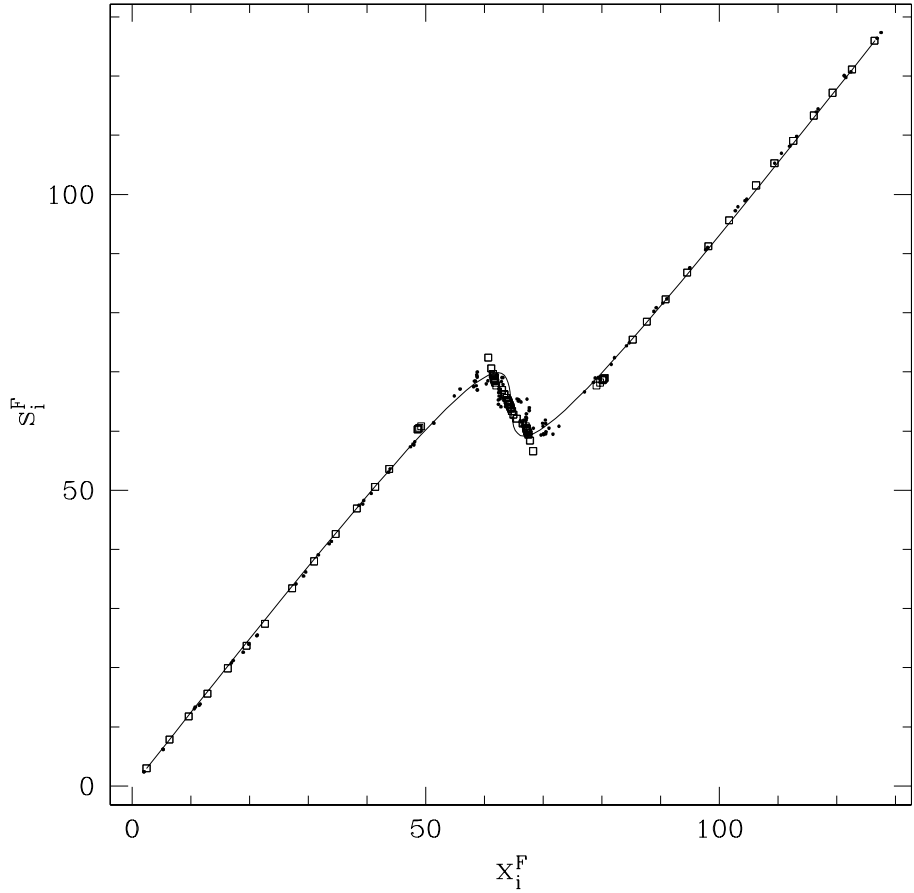,angle=0,height=5in}}
\caption{The real space-redshift space coordinate relationship for a
Zel'dovich pancake.  The solid line shows the relationship as given by
the output of an N-body simulation.  The squares show the relation as
given by the VELMOD-like scheme used to approximate the real-space
density field.  Finally, the solid points show the output from PLA.}
\label{fg:triple}
\end{figure}

In order to illustrate this, we have run a simulation in which a plane
wave density field is laid down along the line of sight.  Its
amplitude is such that there are triple-valued zones at $a=1$.  The
solid lines in Figure~\ref{fg:1d} show the initial and final
density field of this distribution in both real and redshift space.
The solid line in Figure~\ref{fg:triple} demonstrates the
existence of triple value zones.  There, we plot the relationship
between real and redshift particle coordinates for our simulated
Zel'dovich pancake.

Given the ``observed'' redshift space density field, $\delta^F({\bf
s})$ in panel d) of Fig.~\ref{fg:1d}, we have already described
how to generate a set of redshift space coordinates ${\bf s}^F_i$.
Since the simulation is approximately in the Zel'dovich regime, it is
assumed that if we can estimate the real space coordinates, we can use
that information to estimate the position of a particle on the uniform
grid,${\bf q}_i$.  However, given the degeneracy of the mapping from
redshift to real space, this is no simple task.

Willick et al. (1997; also Sigad et al. 1998) suggest a likelihood
approach to breaking this degeneracy, called VELMOD.  Part of VELMOD
relates the observed redshift space density field to a test value of
the real space density field.  We use a similar approach here.  We
first assume that the real space particle field is the one generated
from the assumption of no orbit crossings.  That is,
equations~(\ref{eq:partpos}) is used to approximate the positions of
the particles.  One may then compute a real space density field from
the particle field approximations, and from that the potential field,
$\phi({\bf x})$ can then be computed.  The assumption of linearity
gives the following relation:
\begin{equation}
{\bf p}({\bf x})=-\frac{\nabla\phi({\bf
x})}{(a_0\ddot{D}_0+\dot{a}_0\dot{D}_0)}\ .
\label{eq:p_velmod}
\end{equation}
This makes $\tilde{\bf s}^F$ a straightforward function of ${\bf x}^F$ via 
equation~(\ref{eq:zeld}).  For triply valued $\tilde{\bf s}^F_i$, we
can then determine a posterior probability that the particle is at
real space position, ${\bf x}^F$:
\begin{equation}
P({\bf x}^F_i | \tilde{\bf s}^F_i)\propto [1+\delta^{OLD}({\bf x}_i)]
\delta^D\left[{\bf x}_i-{\bf x}(\tilde{\bf s}_i^F)\right]
\frac{\partial x({\bf s}_i)}{\partial \tilde{s}_i}
\end{equation} 
where $\delta^{OLD}({\bf x})$ is the previous iteration of the
estimated real space density field.  This is simply the discrete form
of the continuous distribution function used in the VELMOD approach.
Using this distribution function, a real space coordinate is randomly
assigned to each triple-valued particle.  The real space density field
and potential are then recomputed and the process is repeated until
satisfactory convergence is reached.

We apply this VELMOD-like approach to our Zel'dovich pancake.  The
dotted and short dashed lines in the lower-left panel of
Figure~\ref{fg:1d} show the initial guess and final estimate of
the real space density field.  While the fit between the true and
estimated real space density fields are good, they are not perfect.
One way of thinking about this is that
equation~(\ref{eq:p_velmod}) assumes that the final velocity of a
particle is linearly proportional to the force on that particle.  In
the Zel'dovich approximation, this is a good assumption.  However, as
the limits of that approximation are approached, the relation
between acceleration and velocity may become more complex.  On scales
on which shell crossings occur, the two may even be of opposite signs.
One approach that people have historically used is to simply smooth
density fields until all structures are linear.  It should be noted
this density field is {\it not} our final estimate of the real-space
field.  Rather, we are using it as a working model to set up redshift
space constraints for PLA, which makes no assumptions about the
linearity of orbits.

This artificial steepening is also apparent in the estimated
relationship between real and redshift space coordinates, as is
illustrated by the open square points in Figure~\ref{fg:triple}.
Notice, however, that we have qualitatively reproduced the features of
the tripled value zones.

At the final iteration of VELMOD, not only are assumed final positions
of the particles computed, but so are their velocity/displacement
vectors, ${\bf p}_i$.  If linear theory approximately holds, then each
of those particles ought to have originated at:
\begin{equation}
\tilde{q}^\alpha_i\equiv \tilde{s}^{F\alpha}_i-[1+\delta^K_{\alpha
3}f(\Omega_M,\Omega_\Lambda,t_0)]\ p_i^\alpha\ .
\end{equation}

We may now return to the problem posed at the beginning of this
section:  How do we compute the ``best'' permutation matrix, ${\bf
M}_{ij}$?  We find the matrix which minimizes:
\begin{equation}
\chi^2[{\bf M}_{ij}\tilde{\bf q}_j,{\bf q}_i]\equiv 
\sum_{i}\left[{\bf q}_i- {\bf M}_{ij}\tilde{\bf q}_j \right]^2\ .
\end{equation}
In order to actually perform this minimization, we apply a simulated
annealing method (Press et al. 1992) Even for systems with a number of
triple-value zones, the first guess of final and initial particle
position pairings produces rapid convergence.  One can then permute
$\tilde{\bf s}^F_i$ into ${\bf s}^F_i$, resulting in a well-motivated
set of boundary constraints.

\subsection{Computing Trajectories in Redshift Space}

We will now consider the simultaneous determination of the orbits of
interacting particles when the boundary constraints are given in
redshift space.  Let us say that we have an isolated, uniform grid of
particles at $a=0$, with positions given by $\{ {\bf q}_i \}$, and at
$t=t_0$, those particles are ``observed'' at redshift coordinates, $\{
{\bf s}^F_i \}$.  We write down the trajectories of the particles as
the sum of a part given by linear perturbation theory, and
coefficients times basis functions.  However, unlike the discussion of
PLA in real space (GS), redshift space gives us heterogeneous
constraints on our basis functions.  In real space, the basis
functions were constrained such that:
\begin{equation}
f_{n}(t_0)=0 \ \ ; \ \  \lim_{a\rightarrow 0} a^2\dot{f}_n(t)=0
\label{eq:fcons}
\end{equation}
In GS, we showed that these constraints can be satisfied by using
\begin{equation}
f_n=\sum_{m=n}^{m_{max}}b_{nm} D(t)^m\left[D_0-D(t)\right]\ ,
\end{equation}
where $b_{1m}=\delta^{k}_{1m}$, and the higher order coefficients are
based on fitting to an N-body simulation.

In redshift space, however, we need to define a slightly different set
of basis functions, $\tilde{f}_n(t)$.  In this case the basis
functions along the line of sight must satisfy
\begin{equation}
\tilde{f}_n(t_0)+\frac{a_0\dot{\tilde{f}}_n(t_0)}{ \dot{a}_0}=0 \ \ ;
\ \  \lim_{a\rightarrow 0} a^2\dot{\tilde{f}}_n(t)=0\ ,
\label{eq:ftildecons}
\end{equation}
in order that the varying the coefficients of $\tilde{f}_n$ do not
change the corresponding radial redshift space coordinate.

To satisfy these constraints, we introduce a complementary set of
basis functions to those introduced in real space:
\begin{equation}
\tilde{f}_n(t)=f_n(t)-\frac{\dot{f}_n(t_0)}{\dot{a_0}}\frac{a^n}{1+n}\
.
\label{eq:tildedef}
\end{equation}
It can be shown that these functions satisfy
equation~(\ref{eq:ftildecons}) if the real space basis functions
described above are used.  As with the unaccented basis functions,
only the first function goes linearly or slower at early times.  Using
these basis functions, we are able to describe the trajectory of any
particle as:
\begin{eqnarray}
x^{\alpha}_i(t)=x^{(0)\alpha}_i+D(t)(s^{\alpha F}_i-
x^{(0)\alpha}_i)+\sum_n C^{\alpha}_{in}f_n(t) & ;
& \alpha=1,2 \\ 
x^{\alpha}_i(t)=x^{(0)\alpha}_i+\frac{D(t)(s^{\alpha F}_i-
x^{(0)\alpha}_i)}{[1+f(\Omega_M,\Omega_\Lambda,t_0)]}+\sum_n
C^{\alpha}_{in}\tilde{f}_n(t) & ; & \alpha=3 \ ,
\end{eqnarray}
where $x^{(0)\alpha}_i$ are the set of some physically self-consistent
orbits, as output from an N-body code.  Of course, we may also set
$x^{(0)\alpha}_i=q_i$.  

In GS, we showed that by simultaneously minimizing the action, $S$,
for each coefficient, $C^{\alpha}_{i,n}$, then the equations of motion
of the particles are necessarily satisfied, and hence, we may find
these orbits.  That is,
\begin{equation}
\frac{\partial S}{\partial C^{\alpha}_{i,n}}=
\int_0^{t_0} dt \left[f_n(1-\delta^K_{\alpha
3})+\tilde{f}_n\delta^K_{\alpha 3}\right] \left[
-\frac{\partial}{\partial
t}(a^2\dot{x}_i^{(1)\alpha})+
\frac{\partial{\phi}^{(0)}_i}{\partial x_i^{\alpha}}-
\frac{\partial{\phi}_i}{\partial x_i^{\alpha}}
\right]\ ,
\label{eq:Z_PLA}
\end{equation}

We will solve equation~(\ref{eq:Z_PLA}) for all possible basis
functions simultaneously, by determining the coefficients such that
the kernel
\begin{equation}
{\bf g}_i(t)\equiv -\frac{\partial (a^2\dot{\bf x}_i^{(1)})}{\partial
t} +\left( \nabla\phi^{(0)}_i-\nabla\phi_i \right)
\end{equation}
vanishes at all times, and for all particles.

In order to do this, we can use the metric:
\begin{equation}
X^2\equiv\sum_i \int dt |W(t) {\bf g}_i(t)|^2\ ,
\end{equation}
where $W(t)$ is an arbitrary weighting function.  By minimizing $X^2$,
we find the set of trajectories which come closest to satisfying the
equations of motion.

Perturbations of the basis functions may be approximated by:
\begin{equation}
\delta C^\gamma_{i,n}\simeq \left[ \int dt
\Delta^\alpha_{i,m,\beta}(t)\Delta^\alpha_{i,n,\gamma}(t)\right]_{m\gamma,n
\beta }^{-1}\left[\int dt W(t)^2
g^{\alpha(old)}_i(t)\Delta^\alpha_{i,m,\gamma}(t)\right]_{n\beta}\ ,
\end{equation}
where 
\begin{equation}
\Delta^\alpha_{i,n,\beta}(t)\equiv \frac{\partial
g^\alpha_i(t)}{\partial C^\beta_{i,n}}=
-\delta_{\alpha\beta}\frac{\partial [a^2 x_i^{\alpha(1)}
f_n(t)]}{\partial t}-\frac{\partial^2 \phi_i}{\partial
x_i^\alpha\partial x_i^\beta} f_n(t)\ .
\end{equation}
Where $\beta=3$, the basis functions, $f_n(t)$ should be replaced by
$\tilde{f}_n(t)$.

We illustrate the results of this method on the Zel'dovich pancake in
Figure~\ref{fg:1d} and Figure~\ref{fg:triple}.  For this, we
have used 4 basis functions and two iterations.  For each iteration,
we took the constraint pairs, $\{ {\bf q}_i,{\bf s}^F_i \}$, and used
PLA to compute the best fit full trajectory.  We then evaluated the
positions and velocities of the particles at $a\simeq 0.01$, and ran
the particles through the PM code again.  

The long dashed lines in panels c) and d) of Figure~\ref{fg:1d}
represent the density field of the second iteration in redshift and
real space, respectively.  While by the nature of the constraints we
would necessarily expect the redshift space density field to converge
to the ``true'' field, we have no such guarantee in real space.
Nevertheless, the real space density field does seem to give a
somewhat better fit than the initial estimate given by the VELMOD-like
approach, especially around the edge of the peak, PLA gives a smoother
edge.  

Additionally, even though the peak is nonlinear in the sense that
$\delta > 1$ in real space, PLA is able to very successfully generate
an initial density field.  Figure~\ref{fg:1d}a) and b) show the
redshift and real space density field determined by PLA, as well as
the true initial density field.  PLA is able to determine the
amplitude of the initial peak to about $10\%$.  However, it should be
noted that PLA may erroneously generate too much small scale power.
Given some {\it a priori} knowledge of the power spectrum, however,
one may use a power-preserving filter like the one described in GS in
order to appropriately smooth the field.  In this 1-d case, in which
we only expect a single mode, using such a filter would be gratuitous.

Finally, in Figure~\ref{fg:triple}, we show the relationship
between real and redshift space coordinates as determined from the
output of the PLA code.  Note that the PLA points turn over more
smoothly than our initial guess points.  This is due to the fact that
PLA assumes the field to be evolving from an initially uniform
particle field, while the VELMOD-like approach makes no such
assumption.

\section{A High Resolution Test of the Code}

To illustrate PLA's success as a reconstruction scheme somewhat into
the nonlinear regime, we have run a high resolution simulation, with
$N_p=128^2$, $N_g=256^3$, and with $L=800 h^{-1}$ Mpc.  The cosmology
used is $\Omega_M=0.3$, $\Omega_\Lambda=0.7$, and $\sigma_8=0.7$.  We
then apply PLA under the assumption that the cosmology was known to
reconstruct the field.

\begin{figure}
\centerline{\psfig{figure=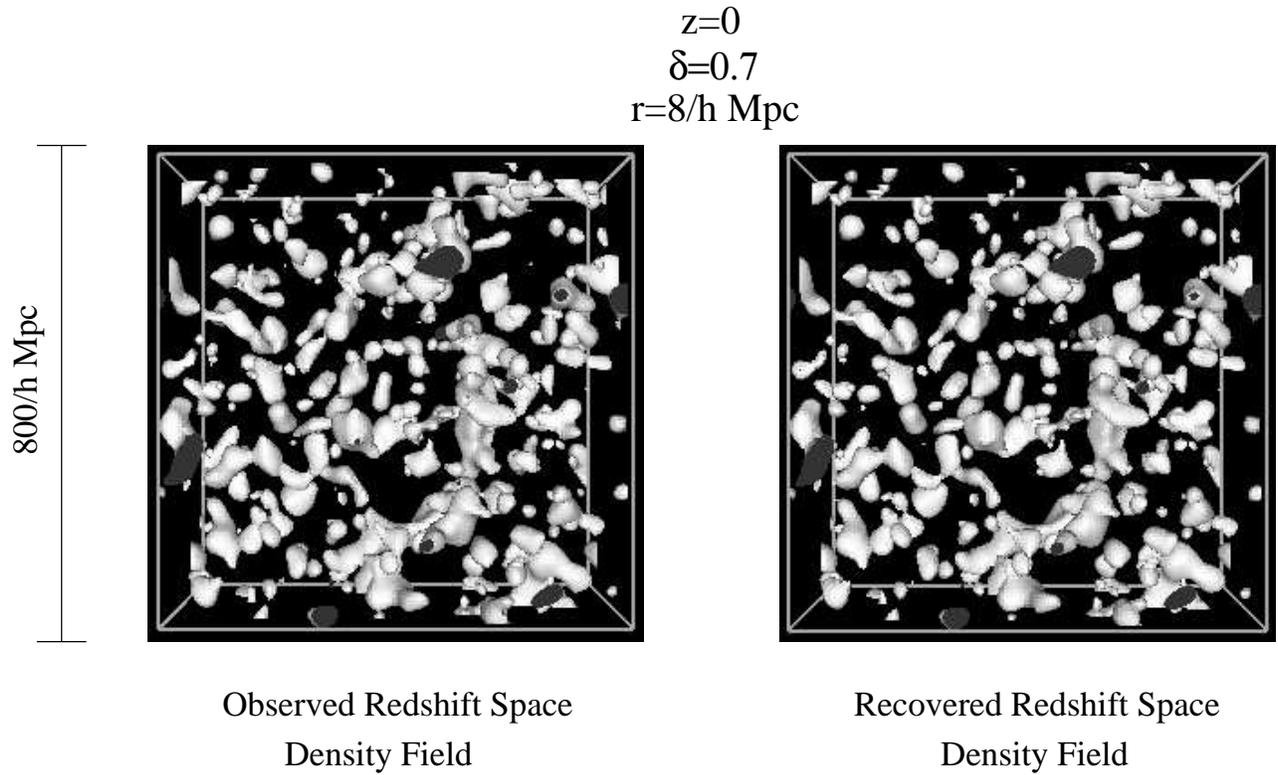,angle=-90,height=4in}}
\caption{A comparison of the smoothed redshift space density fields of
the output of the resolution simulation described in the text, and the
reconstructed field based on those observations.  A contour of
$\delta=0.7$, smoothed with a Gaussian filter of $r=8h^{-1}$ Mpc is
shown.  Note the high level of agreement between structures in the
true and reconstructed fields.}
\label{fg:compare}
\end{figure}

In Figure~\ref{fg:compare} we show a density contour of
$\delta=0.7$ for the observed redshift space density field, smoothed
with a Gaussian filter with a radius of $8 h^{-1}$ Mpc, and a similar
plot, but for the field resulting from our PLA reconstruction.  Recall
that the underlying particle field for the latter is based on running
the reconstructed initial conditions through a PM code and taking the
smoothed density field.  A visual inspection demonstrates that the two
fields are virtually identical.

\begin{figure}
\centerline{\psfig{figure=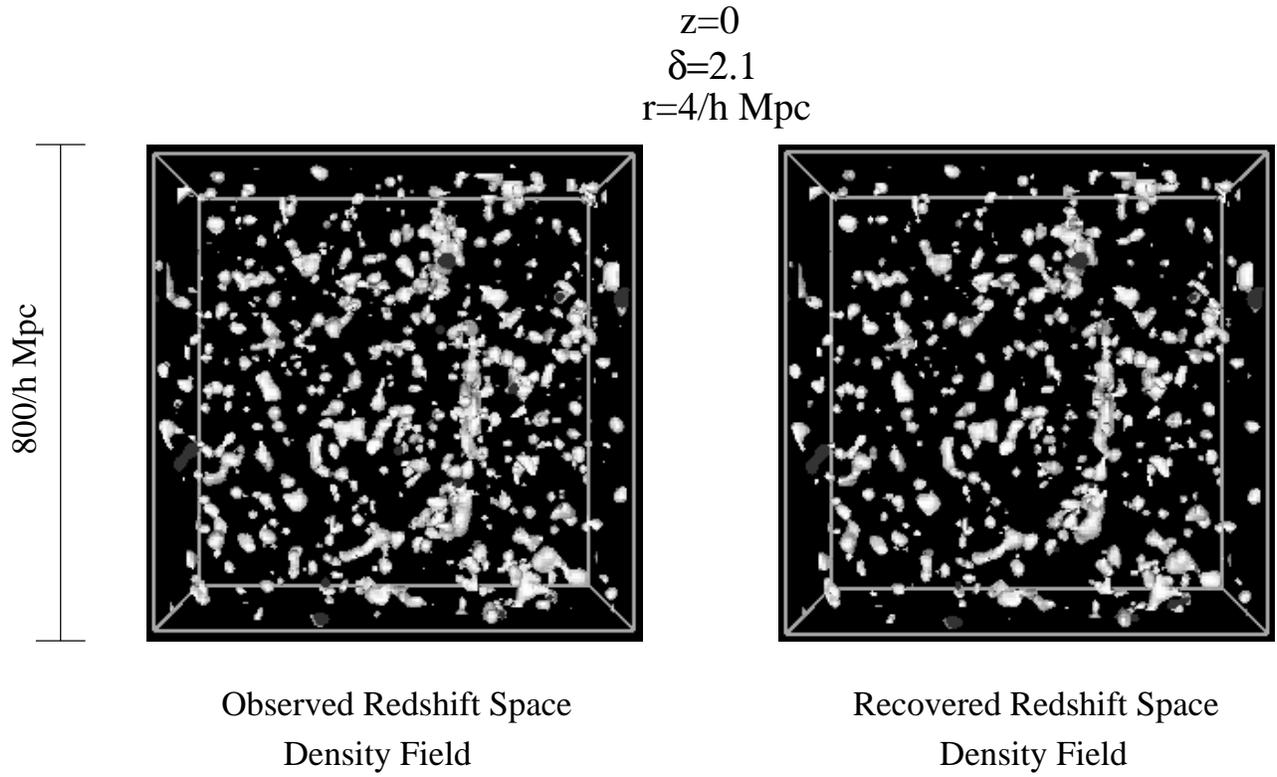,angle=-90,height=4in}}
\caption{As in the previous figure, but with a Gaussian smoothing
radius of $4 h^{-1}$ Mpc, and a density contour of $\delta=2.1$.}
\label{fg:compare2}
\end{figure}

We plot a similar comparison in Figure~\ref{fg:compare2}, except with
a smoothing radius of only $4 h^{-1}$ Mpc, and a density contour of
$\delta=2.1$.  On these scales, too, the fields are very similar.  The
reconstructed field, however, seems to differ somewhat on small
scales.  If we had an {\it a priori} model of the power spectrum, this
small scale power could be suppressed numerically.

\begin{figure}
\centerline{\psfig{figure=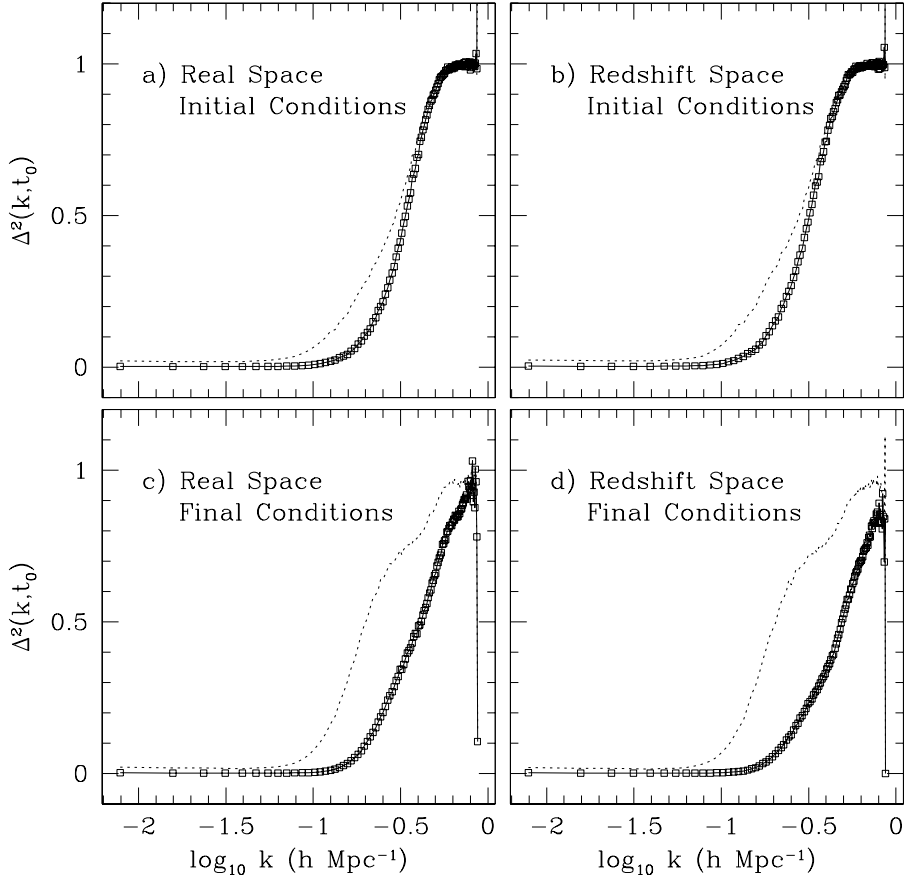,angle=0,height=5in}}
\caption{The Fourier difference statistic for the high resolution
simulation.  This compares the true initial and final conditions in
real and redshift space to those generated by using the reconstruction
scheme using only the observations of the final redshift space density
field.  The dotted line shows the first iteration of PLA, and the
solid line, an additional iteration.  Note that a physical scale of
$16h^{-1}{\rm Mpc}$ (the nonlinear scale) occurs at
$\log_{10}(k)=-0.4$.}
\label{fg:dfit}
\end{figure}

More quantitatively, Figure~\ref{fg:dfit} shows the fit between
the reconstructed and true field as a function of scale.  Narayanan \&
Croft (1999) provide the goodness of fit metric:
\begin{equation}
\Delta^2(k,t)=\frac{\sum [\delta_1({\bf k})-\delta_2({\bf k})]^2}
{\sum [\delta_1({\bf k})^2+\delta_2({\bf k})]^2}\ .
\end{equation}
In this case, $\delta_1({\bf k})$ and $\delta_2({\bf k})$ are the
Fourier transforms of the true and reconstructed redshift density
fields.  For perfect matching on a particular scale, this metric goes
to zero.  For uncorrelated fields, it goes to one.

As Figure~\ref{fg:dfit} illustrates, the fit for all four
comparisons is very good even into the nonlinear regime ($k\simeq 0.4
h\ {\rm Mpc}^{-1}$).  The best fit was found for the final conditions
in redshift space, since this was the actual set of observations to be
matched.  However, it is shown that the real space field at z=0 is
also reconstructed quite well, as are the corresponding initial
conditions.  In particular, the relevant scale is that for which
$\Delta^2=0.5$ for each comparison.  In this case, the real and
redshift initial conditions are well matched down to a scale of $18.6$
and $19.5 h^{-1}$ Mpc, respectively, and the real and redshift space
final conditions are matched down to scales of $15.1$ and $12.7
h^{-1}$ Mpc, respectively.

Another measure of the quality of the reconstruction is the isotropy
of the reconstructed real space density field.  We may do this by
decomposing the power spectrum of the field into Legendre Polynomials
(e.g. Hamilton, 1992):
\begin{equation}
P({\bf k})=\sum_{l\ {\rm even}}{\cal P}_l(\mu)P_{l}(k)\ ,
\label{eq:pldef}
\end{equation}
where $\mu$ is the cosine of the angle between the direction vector,
${\bf k}$, and the line of sight, ${\cal P}_l(\mu)$ are the Legendre
polynomials, $P_l(k)$ are the azimuthally-averaged multipole expansion
of the power spectrum.  Taking the inverse Legendre transform, we find
(Cole, Fisher and Weinberg 1995):
\begin{equation}
P_{l}(k)=\frac{2l+1}{4\pi}\int_{-1}^{1} d\mu \int_0^{2\pi} d\phi
P({\bf k}){\cal P}_l(\mu)\ .
\label{eq:inverse}
\end{equation}
Here, we will only be using the monopole and quadrupole moments,
which, as a reminder are ${\cal P}_0(\mu)=1$ and ${\cal
P}_2(\mu)=(3\mu^2-1)/2$.  We may thus measure the isotropy of the
distribution by computing the ratio:
\begin{equation}
Q_x(k)\equiv \frac{P_2(k)}{P_0(k)}\ .
\end{equation}
Since the real universe is isotropic, an accurately computed
reconstruction should have a quadrupole equal to zero on all scales.

\begin{figure}
\centerline{\psfig{figure=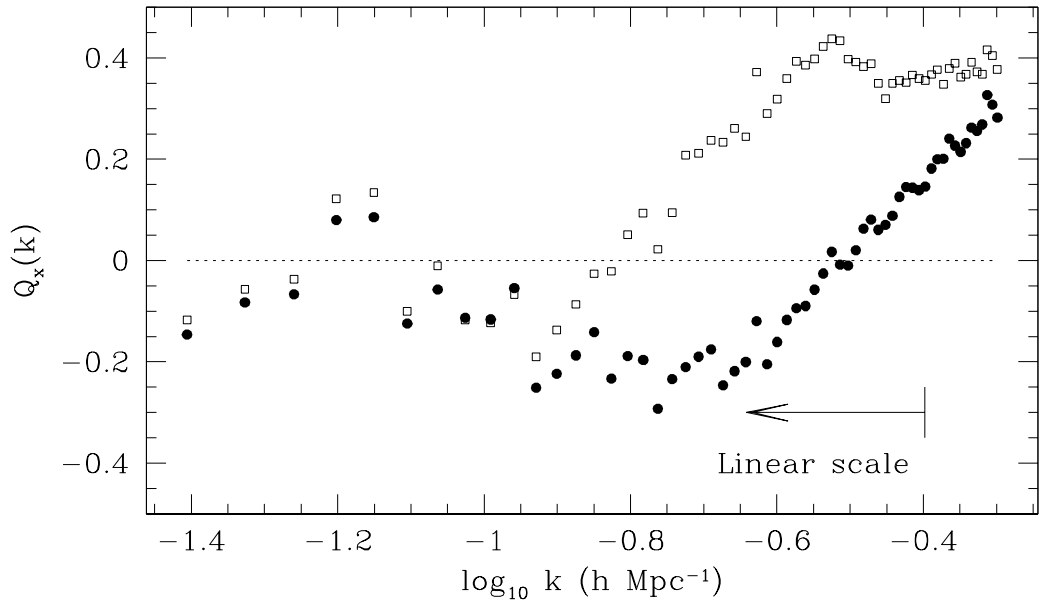,angle=0,height=6in}}
\caption{The quadrupole moment ratio as a function of $|k|$ for the
reconstructed high-resolution real space density field
simulation. Solid squares represent the results from one iteration of
PLA, while solid circles are the results from a second iteration. }
\label{fg:qfit_big}
\end{figure}

Figure~\ref{fg:qfit_big} shows the quadrupole ratio as a function of
scale for both 1 and 2 iterations.  Two things are clear from this
plot, however.  First, the overall isotropy does improve with
additional iterations.  Secondly, there is a systematic effect in
generating these anisotropies which is almost certainly caused by an
anisotropic noise term.  In the appendix, we use the quadrupole and
hexadecipole moments to illustrate that this effect is dominated by
noise, rather than by a systematic underestimate of the radial
velocity term, for example.

This noise term comes out of the reconstruction scheme, itself.  In
the previous section, we discussed how one goes about approximately
rewinding the trajectory of a particle in order to determine its
initial constraint.  However, there was an assumption that linearity
approximately held.  As structure gets more and more nonlinear, this
assumption will fail to hold, and the particle matching technique will
break down.  Future advances in reconstruction methods will have to
take this into account.  It may be possible to apply PLA in an
iterative and statistical way in order better do this matching.

\section{Applications: Breaking the Bias Degeneracy}

\subsection{Motivation}

Though the reconstruction of nonlinear fields is interesting in its
own right, we have begin this investigation into redshift space for
the purpose of finding out something about cosmology.  As a motivation
for this sort of reconstruction analysis of redshift surveys, we
address the $\Omega_M$-$b$ degeneracy in redshift surveys, and discuss
how the degeneracy might be broken without recourse to outside
dynamical estimates of $\Omega_M$.  We will test the effectiveness of
using PLA to break the degeneracy.

We begin by introducing the problem as it appears in the linear
regime.  Excellent recent reviews of this topic is given by Hamilton
(1998) and by Strauss \& Willick (1995), and we will therefore present
only an overview of the linear biasing problem.

Let us first suppose that we have the full velocity field
information about a group of almost uniformly distributed particles.
Let us further suppose that these particles are biased with
respect to some true underlying CDM field, such that:
\begin{equation}
\delta_g({\bf x},t_0)=b\delta({\bf x},t_0)\ ,
\label{eq:delg}
\end{equation}
where the unsubscripted $\delta$ is the CDM density field, and
$\delta_g$ (for galaxies) represents the density field of some biased
tracer of the mass.  Note that a straight linear biasing model may be
replaced with any deterministic, local, and monotonically increasing
function of $\delta({\bf x},t_0)$ without changing the essence of the
discussion or the biasing problem in general.  Blanton (1999) provides
an excellent review of various types of biasing models.

With linear bias greater than unity the velocity field is of lower
amplitude than one would directly infer from a measurement of
$\delta_g$.  From (\ref{eq:delg}), equation~(\ref{eq:lin}) may
be replaced by:
\begin{equation}
\nabla\cdot{\bf v}({\bf x},t_0)=-\beta\delta_g({\bf x},t_0)\ ,
\end{equation}
where 
\begin{equation}
\beta\equiv \frac{f(\Omega_M,\Omega_\Lambda,t_0)}{b}\simeq 
\frac{\Omega_M^{0.6}}{b}
\end{equation}
and 
\begin{equation}
{\bf v}=f(\Omega_M,\Omega_\Lambda,t_0)\ {\bf p}\ .
\end{equation}
Thus, the divergence of the velocity field and density field are
related by the same proportionality constants for all combinations of
$\Omega_M$ and $b$ which yield the same $\beta$.  This is the crux of
the degeneracy problem.  

In a redshift survey, we do not actually know the divergence of the
velocity field, but rather must infer it through anisotropies in the
redshift density field.  Kaiser (1987) shows that in the linear
regime, there is a straightforward relationship between the real and
redshift space density fields in k-space in the d.o.a.:
\begin{equation}
\delta^s_g({\bf k})=\hat{\bf S}\delta^x=\delta^x({\bf k})[1+\beta \mu^2] \ ,
\end{equation}
where $\hat{\bf S}$ is the linear redshift distortion operator,
$\delta^s_g({\bf k})$ is the Fourier transform of the galaxy density
field in redshift space, and $\delta^x({\bf k})$ is the Fourier
transform of the underlying CDM density field in real space.

From this relationship, a redshift space power spectrum field may be
compared to the real space power spectrum:
\begin{equation}
P_g^s({\bf k})\equiv |\delta^s_g({\bf k})|^2
=P({\bf k})[1+\beta\mu^2]^2\ .
\end{equation}
We can further decompose the power spectrum field into Legendre
polynomials (equation~\ref{eq:pldef}; Hamilton 1992).  We then
take the inverse Legendre transform (equation~\ref{eq:inverse};
Cole, Fisher and Weinberg 1995) and compute the quadrupole moment.

Taking the Legendre expansion of the form of the linear density field
(equation~\ref{eq:pldef}) above, we find that the moments of the
biased, redshift space distribution may be related to the underlying
real space distribution as follows (Hamilton 1998):
\begin{eqnarray}
P^s_{g,0}(k)&=&\left(1+\frac{2}{3}\beta+\frac{1}{5}\beta^2\right)P(k)
\\ 
P^s_{g,2}(k)&=&\left(\frac{4}{3}\beta+\frac{4}{7}\beta^2\right)P(k) \ .
\end{eqnarray}
Though the underlying power spectrum is not known, this quadrupole
ratio may be computed as:
\begin{equation}
Q_s(k)\equiv\frac{P^s_{g,2}(k)}{P^s_{g,0}(k)}=
\frac{\frac{4}{3}\beta+\frac{4}{7}\beta^2}
{1+\frac{2}{3}\beta+\frac{1}{5}\beta^2} \ .
\end{equation}
Since the multipole expansion may be computed directly from the
observed redshift space density field, it is clear that in the linear
regime, the quadrupole ratio is degenerate for a particular value of
$\beta$.

In addition, $\beta$ may be estimated through the use of
distance-velocity comparisons.  In practice, measurements of $\beta$
are still quite difficult due to the noisy data involved.  Willick
(2000) gives a review of current estimates, and finds a value of
$\beta=0.5\pm 0.04$ from the IRAS velocity field and the Mark III
Catalog used to estimate distances.  This is relatively unchanged from
the earlier estimate based on measured redshift-space anisotropy by
Cole, Fisher, \& Weinberg (1995) of $\beta=0.54\pm 0.3$.  Ballinger et
al. (2000) estimate $\beta=0.4\pm 0.1$ for the IRAS 0.6 Jy PSCz survey
from analysis of anisotropies.  Similar values are found from the
Optical Redshift Survey catalog (Baker et al. 1998).

After estimating $\beta$, how does one estimate $\Omega_M$ without
recourse to direct mass estimates?  We answer this by first pointing
out that a correctly reconstructed real space density field will be
perfectly isotropic.  That is, $Q_x(k)=0$ on all scales.
Nonlinearities is the evolution of the density field will mean that
the linear redshift distortion operator will no longer be valid on all
scales.  By reconstructing fields using PLA for different combinations
of $\Omega_M$ and $b$, and measuring the quadrupole moment for the
reconstructed field, we can determine the ``true'' cosmology as that
which minimizes the anisotropy.

\subsection{Simulations}

\subsubsection{Constraining Bias}

In order to test this approach, we have run two simulations, each
extending only into the mildly nonlinear regime.  We have found that
simulations which contain highly nonlinear structure do not
effectively differentiate between different cosmologies due to the
excessive noise and difficulty of doing the particle matching on small
scales.

\begin{figure}
\vskip -3.0in
\centerline{\psfig{figure=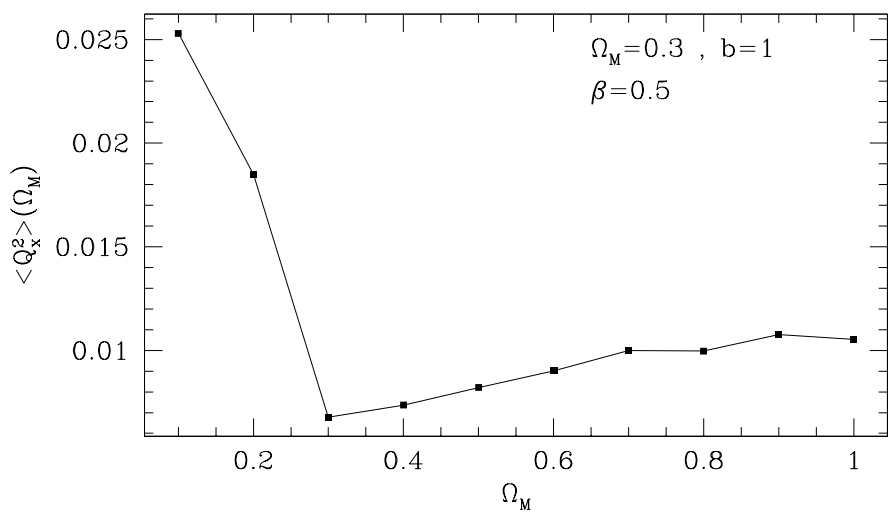,angle=0,height=5in}}
\caption{The weighted-average real space quadrupole ratio for the
reconstructed density fields with various assumed cosmologies in the
$\Omega_M=0.3$ simulation.  Note that the quadrupole ratio is minimized
for $\Omega_M=0.3$, indicating that PLA reconstruction effectively
breaks the $\beta$ degeneracy.}
\label{fg:qfit_om03}
\end{figure}

\begin{figure}
\vskip -3.0in
\centerline{\psfig{figure=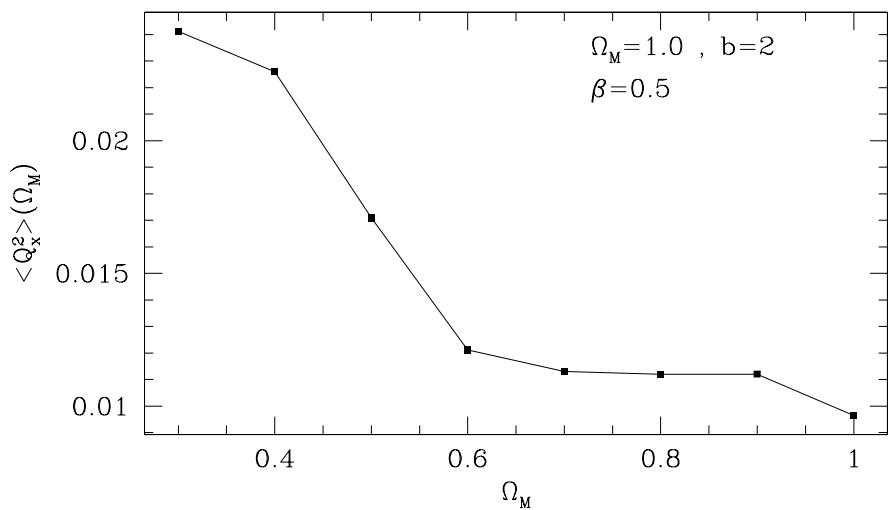,angle=0,height=5in}}
\caption{As in the previous figure, but for the $\Omega_M=1$
simulation.  Here, the quadrupole ratio is minimized for $\Omega_M=1$,
again indicating the reconstruction method has broken the degeneracy.}
\label{fg:qfit_om1}
\end{figure}

The two simulations were each run with $\beta=0.5$ and a boxsize of
$L=1000 h^{-1}$ Mpc, the first with $\Omega_M=0.3$,
$\Omega_\Lambda=0.7$, and $b=1$, and the second with $\Omega_M=1$,
$\Omega_\Lambda=0$, and $b=2$.  Each simulation was run with $64^3$
particles, and $128^3$ gridcells, and the PLA reconstruction used 4
basis functions.  The ``observations'' of these simulations were the
redshift-space density fields in the d.o.a.  After reconstructing the
initial density field for a particular assumed bias, we ran the
initial conditions through a PM code, and measured the quadrupole
moment in real-space, where it should vanish.  The ``mean'' quadrupole
moment is estimated as:
\begin{equation}
\langle Q_x^2 \rangle=\frac{\int dk \ Q^2_x(k)\ k^2\ e^{-k/k_l}}{\int dk\ k^2\
e^{-k/k_l}}\ ,
\end{equation}
where $k_l$ is a limiting scale at which point grid and/or nonlinear
effects will become important.  We have assumed $k_l=20 \times
2\pi/L_{box}$, but found similar results for $k_l=\infty$.  The peak
of this contribution occurs around $3k_l$, or on a physical scale of
$\sim 16 h^{-1}$ Mpc, the nonlinear scale.  The results of each model
tested are shown in Figure~\ref{fg:qfit_om03} for the
$\Omega_M=0.3$ simulation, and Figure~\ref{fg:qfit_om1} for the
$\Omega_M=1$ simulation.

In each case, we find that the best fit value of $\Omega_M$
corresponds to the actual value of $\Omega_M$ used in the simulation.
That is, by tracing the detailed evolution of a mildly nonlinear
field, PLA can effectively break the $\beta$ degeneracy.

In estimating this effect from a real survey, we would have to Monte
Carlo observations based on the survey geometry, selection function,
and the like, in order to estimate $\Omega_M$ and its errors.  While
direct error estimation is difficult with only two realizations, the
results are quite suggestive that this will be an effective way to
constrain $\Omega_M$ directly.  Susperregi (2000) uses an Eulerian
least action code to similarly show that the bias degeneracy may be
broken through accurate reconstruction.

\begin{figure}
\centerline{\psfig{figure=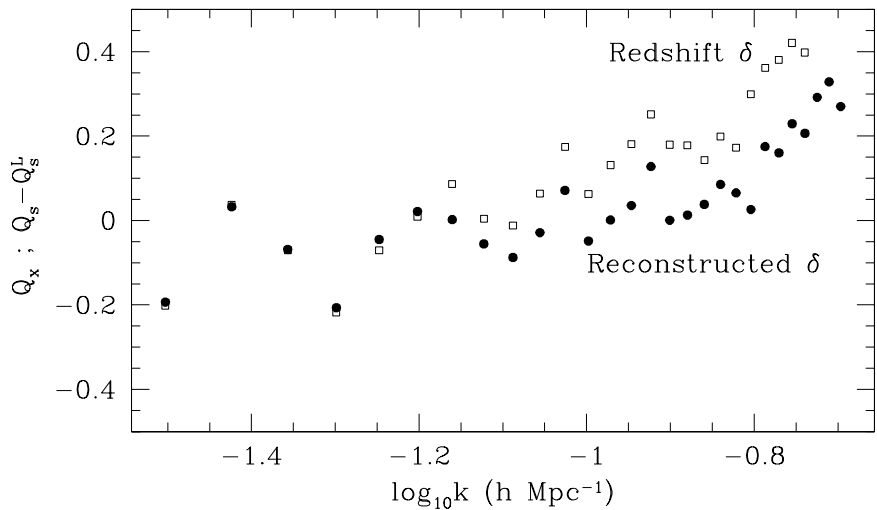,angle=0,height=5in}}
\caption{A comparison of the scale dependent quadrupole moment ratio
for the reconstructed real space density (filled circles), and the
``observed'' redshift space density (open squares), with the linear
quadrupole term subtracted.  Each estimate was done using the
$\Omega_M=0.3$ low resolution simulation.  Note that while both
statistics are approximately zero at large scales, there is a
systematic divergence at small scales.  This turnoff occurs at larger
scales for the direct redshift estimate than for the estimate based on
the reconstructed field.}
\label{fg:redreal}
\end{figure}

As a final test of PLA as a redshift space reconstruction scheme, we
show that one may obtain somewhat better estimates of $\beta$, itself,
from the assumed isotropy of the reconstructed real space density
field than from the redshift space anisotropy.  This is illustrated in
Figure~\ref{fg:redreal}, in which we show a comparison between
$Q_x(k)$ of the reconstructed field for simulation 1, assuming the
correct value of $\Omega_M=0.3$, and the corresponding quadrupole
moment residuals, $Q_s(k)-Q_s^L$ of the observed redshift space field.
Note that the term $Q_s^L$ is simply that which one would estimate
from an assumption of the correct value of $\beta$.

On large scales, these two statistics are almost identical.  However,
on smaller scales, when nonlinearities begin to become important, the
two statistics both exhibit anisotropies.  The redshift space field,
however, becomes anisotropic on larger scales.  Since the
corresponding uncertainties in $\beta$ are approximately proportional
to one over the square root of the number of modes probed, we may
relate the expected uncertainties from the reconstructed field to that
from the redshift field as:
\begin{equation}
\frac{\sigma_\beta^{PLA}}{\sigma_\beta^{Z}}\simeq 
\frac{\sigma_Q^{PLA}\frac{\partial \beta}{\partial Q^{PLA}}(k^{PLA}_{{\rm max}})^{-3/2}}
{\sigma_Q^{Z}\frac{\partial \beta}{\partial Q^{Z}}(k^{Z}_{{\rm
max}})^{-3/2}}\ ,
\end{equation}
where the superscript ``Z'' refers to the estimate from the redshift
space field and the superscript ``PLA'' refers to the estimate from
the reconstructed real space field.  For $\Omega_M=0.3$, all of the
partial derivatives are almost exactly one.  Moreover, estimates of
the scatter in the quadrupole estimates show that
$\sigma_\beta^{PLA}\simeq 0.9 \sigma_\beta^{Z}$.  Finally, since
Figure~\ref{fg:redreal} shows $k^{PLA}_{{\rm max}}\simeq 1.25
k^{Z}_{{\rm max}}$, the approximate relation between the uncertainty
in the bias between the two errors is $\sigma_\beta^{PLA}\simeq 0.65
\sigma_\beta^{Z}$.  Thus, we are able to generate a somewhat better
constrained estimate of $\beta$ from the reconstructed density field,
as compared to the observed field.

\section{Future Prospects}

This paper has discussed the problem of reconstructing the underlying
real space CDM density field and its evolution from galaxy redshift
surveys under rather idealized conditions.  We have assumed full
sampling, a constant linear bias relation with no morphological
segregation, the distant observer approximation, no errors in
measurement, and a very regular geometry.  We have shown that
theoretically even a mildly nonlinear field can be reconstructed using
PLA to break the bias degeneracy.  In actual observations almost none
of these assumptions will hold.  We would like to end this work with a
brief discussion of how these effects might be appropriately modeled,
and mention a few possible candidates of real surveys to which the PLA
method might be applied.

Throughout, we have assumed a cubic geometry.  In simulating a
realistic survey, a mask must be applied such that statistics may be
correctly computed for the true survey volume.  Additionally, for many
observational samples of interest, the distant observer approximation
no longer describes the system adequately.  Over the course of its
lifetime, a galaxy may have traversed a significant angle in the sky,
making the use of Cartesian coordinates difficult.  An obvious
solution to this problem is to write the PLA equations in spherical
coordinates.  

An additional concern in the application of PLA to realistic
observations is that the biasing model that we have used here is a
strictly linear one, and observational evidence suggest that biasing
may be much more complex (e.g. Blanton 1999 and references therein).
In fact, we have only used linear bias in this work for its
simplicity.  PLA would be applicable to any model of deterministic
bias, even one which had an explicit model of morphological
segregation.  It is not clear how one might effectively reconstruct a
field under the assumption of a significantly stochastic bias model.

Even under the simplest of approximations, realistic surveys are not
cubic, not necessarily contiguous, and generally have nonuniform
selection functions.  The issues of contiguity and geometry are
related, in that in both we need to approximate the density field
outside the survey volume in order to correctly estimate the potential
field within.  

Lahav et al. (1994) apply one such technique to the IRAS 1.2 Jy survey
(Strauss et al. 1992; Fisher et al. 1995) in which expansion of the
density field in spherical harmonics is used to reconstruct the field
outside the survey volume.  In essence, this is very similar to
assuming a particular autocorrelation function, and generating an
outside field based on a truncated form of that function and the
observed field near the edges.  In many respects, this is quite
similar to the sort of reconstruction done using the ``Constrained
Initial Conditions'' technique by Hoffman \& Ribak (1991,1992), since
both use observed the observed autocorrelation function to build
realistic external fields around observed structure.

The final issue in real surveys concerns non-uniformity and noise
within the survey volume.  However, for flux limited surveys such as
the IRAS 1.2 Jy and PSCz (Saunder et al. 2000) surveys, at large
distances, shot noise begins to dominate calculations of the density
field.  Since the observed density field is derived from an incomplete
sampling of a finite number of discrete points, the uncertainties in
the observed density field behaves like a Poisson statistic, with
$\sigma\propto 1/\sqrt{N_{gal}}$.  In order to correctly anticipate
the effects of shot noise when running simulations of a particular
survey, a random component needs to be added to the observed density
field.

Moreover, a non-uniform selection function, coupled with observations
in redshift space results in Malmquist bias.  That is, the ``true''
selection function is based on observed fluxes, and hence is a
function in real space.  The observations, however, are of densities
in redshift space.  Strauss \& Willick (1995) give an excellent review
of how these issues are dealt with in reconstructing density fields.

These considerations are all with an eye toward learning about
cosmology from redshift surveys.  For example, the IRAS 0.6 Jy PSCz
Survey (Saunders et al. 2000) is an especially promising recent
candidate for analysis, as it is publicly available and has nearly
full sky coverage.  Hamilton, Tegmark, \& Padmanabhan (2000) have
already estimated $\beta=0.41^{+.13}_{-.12}$ from this survey, and
Ballinger et al. (2000) find a similar result of $\beta=0.4\pm 0.1$.
Both methods used only linear theory, however.  Valentine, Saunders,
\& Taylor (2000) do a somewhat higher order reconstruction, by using
the PIZA method, and find a best fit to the survey with a slightly
higher value of $\beta\simeq 0.5$.  In testing our code, we have found
that one can get a better measure of $\beta$ by using PLA to
reconstruct a density field under some fiducial cosmology, and
comparing cosmologies to see which produce the minimum anisotropy in
the real space field.  We have showed that ideally, PLA can be used to
discriminate between ``degenerate'' pairs of bias and $\Omega_M$.
Finally, we showed that PLA produces a means by which uncertainties in
the measurement of $\beta$ itself can be reduced.

Another interesting prospect is the application to PLA to large
distance redshift surveys, since one of the byproducts of PLA is the
real space density field.  Nusser et al. (2000) use the $D_n-\sigma$
relation of the ENEAR redshift-distance survey (da Costa et al. 2000)
as test particles within the PSCz survey, much as we would wish to do
using PLA.  A reconstruction was done using the method described by
Nusser \& Davis (1995).  Predicted distances from the reconstructed
field can then be compared with the estimated distances from the ENEAR
survey.  They also estimated $\beta\simeq 0.5$.

While the PSCz survey contains $\sim 15,000$ redshifts, the current
generation of redshift surveys is producing an even greater
opportunity to measure statistical and global properties of the
universe.  When complete, the SDSS redshift survey (York et al. 2000),
will produce $\sim 10^6$ galaxy redshifts, and will cover a quarter of
the sky out to Petrosian magnitude, $r_p'=17.7$ to $z\simeq 0.15$.
The 2dF survey (Colless 1999) will ultimately measure redshifts over a
quarter of a million galaxies, out to $b_J=19.5$.

Reconstruction of fields from these enormous datasets will prove a
significant computational challenge.  However, it is well worth it, as
PLA can yield insight into the underlying power spectrum, bias, and
cosmological parameters.

\acknowledgments

I would like to gratefully acknowledge many helpful comments by David
Spergel, Michael Strauss, Vijay Narayanan, and Jim Peebles, as well as
a superior visualization tool by Michael Blanton.  This work was
supported by an NSF Graduate Research Fellowship and NASA ATP grant
NAG5-7154.

\newpage

\appendix
\section{The Effects of Noise on Multipole Moments}

In \S3, we found that the reconstructed real space density field of
the high-resolution simulation had a significant and seemingly
systematic quadrupole moment at small scales.  The question which
arises out of this is, does this quadrupole represent a systematic
error in the reconstruction along the line of sight (such as would
occur, for example, if the assumed value of $\beta$ were incorrect),
or does it represent a stochastic term?

To examine this question, let us consider the following form of a
reconstructed field with a very simple error term:
\begin{equation}
\tilde\delta({\bf k})=\delta({\bf k})\left[ 1+\mu^2 B({\bf k})+\mu^2
\epsilon({\bf k})\right]\ ,
\end{equation}
where ${\bf k}$ is a Fourier space component, $\delta({\bf k})$ is the
true real space field we are attempting to reconstruct,
$\tilde\delta({\bf k})$ is the reconstructed form of the field,
$\mu^2$ is the cosine of the angle between the ${\bf k}$ and the line
of sight, $B$ is a systematic, ``bias'' term, and $\epsilon({\bf
k^2})$ is an anisotropic random error drawn from a $N(0,\sigma^2({\bf
k}))$ distribution.  

If the corresponding real space density field of redshift space
observations have been perfectly reconstructed and contain no noise,
the reconstructed field is simply equal to the true underlying field,
and thus will be perfectly isotropic in $k$-space.  However, let us
imagine that a field contains no noise, but the reconstruction is such
that we have assumed that $\tilde{\delta}({\bf k})=\delta_s({\bf k})$,
or the real space density field is the same as the redshift space
field.  Under those circumstances, $B({\bf k})=\beta$, hence our
terminology.

Finally, let us consider a more general case, one in which we wish to
test for a systematic form of $B({\bf k})$ and for the existence of a
random noise component.  In that case, the reconstructed
three-dimensional power spectrum may be written as:
\begin{equation}
\left< \frac{\tilde{P}({\bf k})}{P({\bf k})}\right>=
1+2\mu^2 B({\bf k})^2 +\mu^4 B({\bf k})^2+\mu^4\sigma^2({\bf k})
\end{equation}

If we then decompose these terms into multipole moments (see \S 5.4.1)
and assume that the noise and systematic terms are simply scale
dependent, we find:
\begin{eqnarray}
\tilde{P}_0(k)&=&P(k)\left[ 1+\frac{2}{3}B(k)+\frac{1}{5}B(k)^2
+\frac{1}{5}\sigma^2(k)\right] \\ 
\tilde{P}_2(k)&=&P(k)\left[ \frac{4}{3}B(k)+\frac{4}{7}B(k)^2
+\frac{4}{7}\sigma^2(k)\right] \\ 
\tilde{P}_4(k)&=&P(k)\left[ \frac{8}{35}B(k)^2+\frac{8}{35}\sigma^2(k)\right]
\end{eqnarray}
We may then look at the relationship between the quadrupole ratio,
$\tilde{Q}(k)\equiv \tilde{P}_2(k)/\tilde{P}_0(k)$ and the hexadecipole
ratio, $\tilde{H}(k)\equiv \tilde{P}_4(k)/\tilde{P}_0(k)$.  We have
two extremes.  In the case of no anistropic noise component, each
ratio is simply a parametric function of $B(k)$, and thus, a
straightforward relation between the two may be plotted.

\begin{figure}
\centerline{\psfig{figure=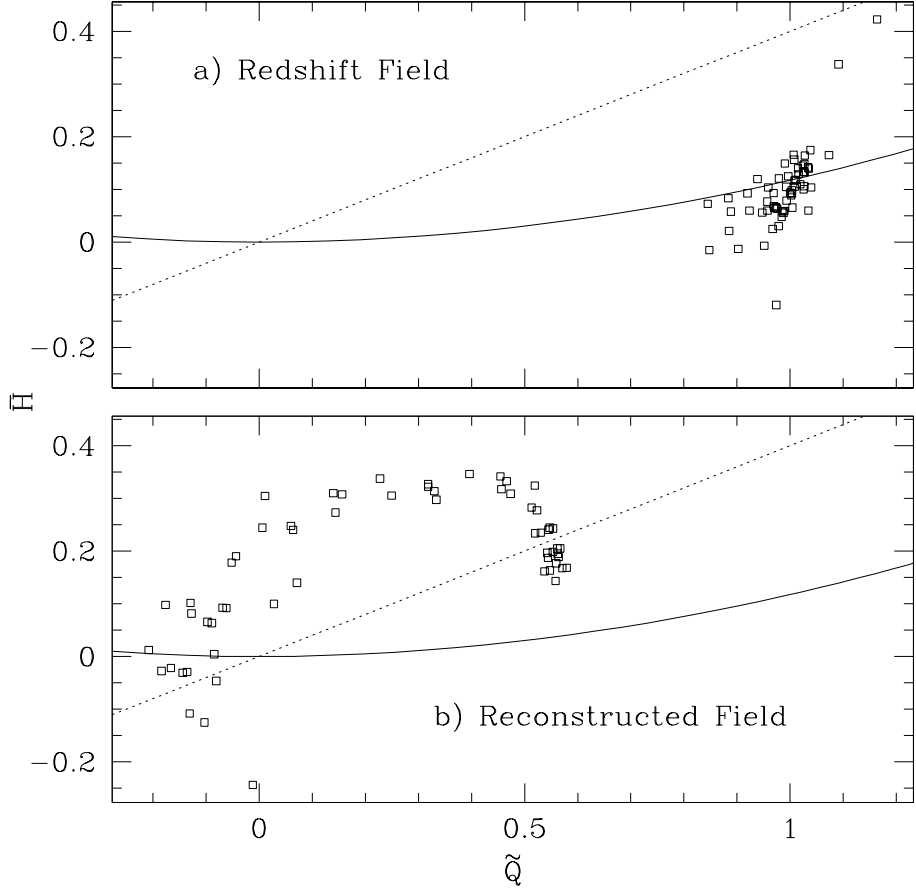,angle=0,height=5in}}
\caption{Scatter plots relating the quadrupole and hexadicpole ratios
in the high-resolution simulations discussion in \S3.  The solid
line shows the expected relation for a systematic anisotropy (the
$\beta$-like term), while the dashed line shows the expected relation
for our simplified anisotropic noise model.  Panel a) shows the ratios
as observed in the redshift space density field.  Note that the
systematic effect dominates, since the anisotropies are completely due
to redshift space distortions.  Panel b) shows the ratios for the
reconstructed field.  The hexadecipole ratio is much higher than what
would be expected from a deterministic effect. }
\label{app:fg:noise}
\end{figure}

If, on the other hand, the anisotropic noise term dominates, we find
the relation:
\begin{equation}
\tilde{H}^{NOISE}(k)=\frac{2}{5}\tilde{Q}^{NOISE}(k)\ .
\end{equation}
We illustrate this in Figure~\ref{app:fg:noise}.  We show that for an
observed redshift density field, the deterministic bias term
dominates.  Though we do not necessarily have the correct form of the
anisotropic error term, this error which would seem to be at the root
of the systematic small scale quadrupole in the high-resolution
simulation.  

Since we may take as a prior that the universe is inherently
isotropic, regularization or iterative techniques might be employed in
future reconstruction schemes which explicitly find a fully isotropic
real space solution.

\newpage

\end{document}